# Widely Spaced Planes of Magnetic Dimers in the $Ba_6Y_2Rh_2Ti_2O_{17-\delta}$ Hexagonal Perovskite

Loi T. Nguyen[1], Daniel B. Straus[1], Q. Zhang[2] and R. J. Cava[1]

[1]Department of Chemistry, Princeton University, Princeton, New Jersey 08544, USA

[2]Neutron Scattering Division, Oak Ridge National Laboratory, Oak Ridge, Tennessee 37831, USA

**ABSTRACT**

We report the synthesis and initial characterization of $Ba_6Y_2Rh_2Ti_2O_{17-\delta}$, a previously unreported material with a hexagonal symmetry structure. Face-sharing $RhO_6$ octahedra form triangular planes of $Rh_2O_9$ dimers that are widely separated in the perpendicular direction. The material displays a small effective magnetic moment, due to the Rh ions present, and a negative Curie-Weiss temperature. The charge transport and optical band gaps are very similar, near 0.16 eV. A large upturn in the heat capacity at temperatures below 1 K, suppressed by applied magnetic fields larger than $\mu_0 H = 2$ Tesla, is observed. A large T-linear term in the specific heat ($\gamma$=166 mJ/mol f.u-$K^2$) is seen, although the material is insulating at low temperatures. These results suggest the possibility of a spin liquid ground state in this material.

# INTRODUCTION

Spin liquids display a novel magnetic state in which the magnetic moments remain fluctuating down to temperatures near absolute zero[1–6]. Invoked as the basis of high temperature superconductivity in the cuprates[7], in quantum spin liquids (QSLs) the magnetic wave functions are entangled at low temperatures. QSL candidates are frequently members of the magnetically frustrated family of materials. The absence of conventional magnetic ordering is often displayed by such materials, with a broad continuum of magnetic excitations seen in inelastic neutron scattering spectra at low temperatures, and large upturns in the heat capacity at temperatures in the sub-Kelvin range. QSL candidates have been proposed based on copper, cobalt, ruthenium, iridium, and rare earth ions[1,2,8–20].

Here we report a new kind of QSL candidate material, a triangular-layered rhodium-oxide-based compound that our initial study indicates to display some of the characteristics that have been attributed to spin liquids. This material, $Ba_6Y_2Rh_2Ti_2O_{17-\delta}$, synthesized in air at ambient pressure at 1500°C, is a structurally ordered hexagonal perovskite analogous to $Ba_6Y_2Ti_4O_{17}$[21]. The crystal structure is based on $Rh_2O_9$ dimers, made from two face-sharing $RhO_6$ octahedra, in a layered triangular array. The distance between two Rh ions within the dimer is about 2.52 Å, and between dimers in the triangular plane is about twice as large. The triangular layers of dimers are separated by non-magnetic $YO_6$ octahedra and $TiO_4$ tetrahedra. Hence, magnetic interactions within the $Rh_2O_9$ dimers, in-plane and out-of-plane magnetic interactions in the three-dimensional structure made from stacked triangular layers of dimers should be considered in an explanation of the properties. The material is a semiconductor with charge transport and optical band gaps of about 0.16 eV, effective magnetic moment of about 1.5 µB per formula unit, and a Curie Weiss temperature of about -2.8 K. No signature of magnetic ordering is observed down to 0.35 K, and an anomalous heat capacity is observed at low temperatures. The magnetic moment, which is clearly present, makes this new material a rare but not unique example of a magnetic rhodium oxide [9,22–24]. The heat capacity at sub-Kelvin temperatures follows a power law with an exponent of 0.57, much smaller than the expected value of 3 (for ordinary three-dimensional magnetic or phononic excitations[25,26]). This upturn under zero applied magnetic field is suppressed by applied fields larger than $\mu_0 H$ = 2 Tesla. Only about 1/3 of the magnetic entropy expected for a spin ½ system is recovered on heating from 0.35 K in zero field, suggesting that spin fluctuations contribute a significant amount of magnetic entropy at sub-Kelvin temperatures in this material. Finally, our spin liquid candidate displays a large temperature-linear contribution to the heat capacity, 166 mJ/mol f.u-K$^2$.[1,15,27], the type of contribution generally attributed to the electronic density of states in metallic materials, which cannot be the case in this electrical insulator.

**EXPERIMENTAL**

Modeled after the dielectric material $Ba_6Y_2Ti_4O_{17}$[28], $Ba_6Y_2Rh_2Ti_2O_{17-\delta}$ was synthesized in polycrystalline form by solid-state reaction using $BaCO_3$ (dried in an oven at 120°C for 3 days), $Y_2O_3$ (dried in a furnace at 1000°C overnight), $RhO_2$ and $TiO_2$ (all purchased from Alfa Aesar, 99.9%, 99.99%, 99.9%, and 99.9% purity, respectively) in stoichiometric ratios as starting materials. The reagents were mixed thoroughly, placed in alumina crucibles, and heated in air at 900 °C for 24 hours. The resulting powder was reground, pressed into pellets, and heated in air at 1100, 1300 and 1500°C for 24 hours at each temperature. Using either Rh or $RhO_2$ as a starting material results in the same final product, excluding the possible influence of nonstoichiometry or partial hydration of the rhodium oxide starting material on the final product obtained.

The phase purity and crystal structure identification were initially determined through powder X-ray diffraction (PXRD) using a Bruker D8 Advance Eco with Cu Kα radiation and a

LynxEye-XE detector. The neutron diffraction experiments, used for the crystal structure determination, were conducted on the Time-of-flight (TOF) powder diffractometer (POWGEN, BL-11A) at the Oak Ridge National Laboratory (ORNL). Four grams of $Ba_6Y_2Rh_2Ti_2O_{17-\delta}$ powder was sealed in a vanadium can with helium exchange gas and loaded in the Powgen Sample Changer. The data were collected at 300 K using a neutron wavelength band of 1.15 –10.5 Å with a central wavelength of 2.665 Å. The structure refinement was performed with GSAS-II[28]. The crystal structure drawings were created by using the program VESTA[29].

The Scanning Electron Microscopy (SEM) with Energy Dispersive X-Ray (EDX) characterization was performed by using a XL30 FEG-SEM equipped with an EVEX EDX in-situ Tensile Stage, and a Gatan MiniCL imaging system. The EDX system provides X-ray acquisition characteristics sufficient for obtaining a high-resolution two-dimensional elemental distribution map over the sample surface. Thermogravimetric Analysis (TGA) of the oxygen content of the material was conducted using a TA Instruments SDT Q600 under flowing $Ar:H_2$ (95:5%). Around 10 mg of $Ba_6Y_2Rh_2Ti_2O_{17-\delta}$ was loaded into an alumina pan and heated from 25°C to 1000°C at the rate 0.5°C/hr. The sample was kept at 1000°C for 30 min, and then cooled to room temperature at the rate 5°C/hr. The phase assemblage of the mixture after this reduction process was determined by using powder XRD.

The magnetic susceptibility of $Ba_6Y_2Rh_2Ti_2O_{17-\delta}$ powder was measured in a Quantum Design Physical Property Measurement System (PPMS) DynaCool equipped with a vibrating sample magnetometer. The magnetic susceptibilities between 1.8 and 300 K, defined as M/H, where M is the sample magnetization and H is the applied field, were measured at different applied magnetic fields. For transport measurements, the sample was pressed, sintered at 1500°C, and cut into pieces of approximate size $1.0 \times 2.0 \times 1.0$ mm$^3$. The resistivity was measured by the DC four-contact method in the temperature range 200 to 300 K in the PPMS. Four Pt contact wires were connected to the samples using silver paint. Below 200 K the sample resistance was too high to be accurately characterized. The specific heat was measured from 200 to 1.8 K in the PPMS DynaCool equipped with a heat capacity option, and to temperatures down to 0.35 K by using a $^3$He system in the PPMS.

The infrared absorption spectrum was taken using a Thermo Scientific Nicolet 6700 FT-IR spectrometer using the attenuated total reflection (ATR) method with a diamond ATR accessory. The bandgap was estimated using the Tauc relation $\alpha h\upsilon = A(h\upsilon - E_g)^n$, where $A$ is a constant, $\alpha$ is the absorption coefficient (cm$^{-1}$), $E_g$ is the bandgap, and $n$ is 0.5 for a direct transition (n = 2 for an indirect transition)[30]. Raman data are collected with a Thermo-Fisher DXR Smart Raman instrument using a 780 nm laser.

**RESULTS and DISCUSSION**

**Figure 1a** shows the SEM and EDX images collected on the $Ba_6Y_2Rh_2Ti_2O_{17-\delta}$ powder sample, confirming that all the elements are present in a single substance. While a variety of Rh$^{4+}$ compounds have been synthesized under high pressure, such as $BaRhO_3$, $CaRhO_3$, $SrRu_{1-x}Rh_xO_3$, $Sr_4Rh_3O_{10}$, and $Lu_2Rh_2O_7$[24,31–36], to the best of our knowledge, $Ba_6Y_2Rh_2Ti_2O_{17-\delta}$ appears to be the only $Rh_2O_9$ dimer compound synthesized at ambient pressure. **Figure 1b** shows the TGA data for $Ba_6Y_2Rh_2Ti_2O_{17-\delta}$, reduced under the flow of Ar:H$_2$ (95:5%) on heating from 25 to 1000°C. The product after reduction is a mixture of Rh, $BaTiO_3$ and $Ba_4Y_2O_7$, identified by using powder XRD. The mass loss is calculated to be around 3.5%, implying that the empirical formula is oxygen deficient, at $Ba_6Y_2Rh_2Ti_2O_{17-\delta}$ ($\delta \approx 0.57$).

Powder neutron diffraction data were used to refine the crystal structure. It was found that $Ba_6Y_2Rh_2Ti_2O_{17-\delta}$ crystallizes in a hexagonal structure with the space group $P6_3/mmc$ (No. 194).

Its powder X-ray and neutron diffraction patterns and structural refinements are shown in **Figures 1c-d**. The structure of $Ba_6Y_2Ti_4O_{17}$ was employed as a starting model for the refinement[21] – for our material, the Ti occupies only the layers of tetrahedra while only the Rh are found in the dimers. Of particular interest relevant to the oxygen deficiency are the oxygen site occupancies, to which neutron diffraction is particularly sensitive. Different models for possible oxygen deficiency were tested, and it was found (**Table 1**) that only the O4 position is not fully occupied, with the occupation of 0.81(2), consistent with the oxygen deficiency observed in the TGA analysis. Finding vacant oxygen sites neighboring a tetrahedral site occupied by a transition element makes our material rare but not unique [37,38]. This partially occupied site is not in the vicinity of the Rh ions and thus its impact on the magnetic properties cannot be considered as primary (near neighbor) or secondary (second nearest neighbor); in fact, the deficient O site is the $7^{th}$ nearest neighbor to the closest Rh in a dimer. The structural refinement indicates the absence of Ti/Rh/Y site mixing disorder to within the sensitivity of the method (a few percent). All the Rh's in the crystal structure are equivalent by symmetry.

The crystal structure of $Ba_6Y_2Rh_2Ti_2O_{17-\delta}$ is shown in **Figure 2a**. Of relevance to the observed magnetism, two $RhO_6$ octahedra are face-sharing to form an $Rh_2O_9$ dimer. In one unit cell, there are two $Rh_2O_9$ dimers linked by $YO_6$ octahedra and $TiO_4$ tetrahedra. The formally mixed valence $Rh^{4+}/Rh^{3+}$-based dimers (with a 50:50 ratio of formal $Rh^{3+}$ to $Rh^{4+}$) are arranged in a triangular lattice as shown in **Figure 2b**. The distance between two Rh atoms within a dimer is quite short (2.52 Å) indicating strong interactions between them. A similarly short Rh-Rh bond length has previously been seen in $Sr_5Rh_4O_{12}$[39], which displays Rh–O chains in its crystal structure. There are two kinds of Rh-O bond lengths in the $Rh_2O_9$ dimer in $Ba_6Y_2Rh_2Ti_2O_{17-\delta}$: 2.00 Å to the outer oxygens and 1.97 Å to the oxygens between the Rh ions, consistent with the presence of strong Rh-Rh interactions, which shorten the Rh-O bonds. Thus there are likely three types of dominant magnetic interactions in $Ba_6Y_2Rh_2Ti_2O_{17-\delta}$, strong interactions within the dimers (d = 2.52 Å), dimer-dimer in-plane interactions (d = 5.94 Å), and plane-to-plane interactions of the dimers (d=14.76 Å), as shown in **Figures 2a-c**. The structural parameters determined in the refinement are summarized in **Table 1** and selected bond lengths are listed in **Table 2**.

The temperature-dependent magnetic susceptibility of $Ba_6Y_2Rh_2Ti_2O_{17-\delta}$ measured under the applied magnetic field of 2 kOe is shown in **Figure 3a**. In an applied field of $\mu_0H = 5$ T, however (data not shown in the figure) the low temperature susceptibility appears to be lower, due to curvature in the M vs H behavior at high fields (**Figure 3b**). Defining the susceptibility as $\Delta M/\Delta H$, and making the measurements in a field that is below the onset of significant curvature in M vs. H, is the intrinsic low field susceptibility of this substance. Above 200 K, the magnetic susceptibility is nearly zero, indicating the presence of a paramagnetic temperature independent term that barely compensates for the core diamagnetism of the constituent elements. Curie-Weiss fitting from 10-50 K results in the Curie-Weiss temperature of -2.8 K and an effective moment of 0.76 $\mu_B$/mol-Rh, as seen in the inset of **Figure 3a**. The small negative Curie-Weiss temperature implies the dominance of weak anti-ferromagnetic interactions, and the relatively low moment observed, comparable to that expected in an ionic picture for isolated $Rh^{4+}$ spin ½ plus $Rh^{3+}$ spin 0 ions ($Ti^{4+}$ has no *d* electrons in its valence orbitals and is not magnetic) in the appropriate ratio determined by the oxygen stoichiometry. The FC/ZFC dc susceptibility in an applied field of 100 Oe is shown in the inset in **Figure 3a**. At low field, there is no bifurcation in ZFC/FC susceptibility down to 1.8 K, indicating that there is no spin glass state or structural disorder in $Ba_6Y_2Rh_2Ti_2O_{17-\delta}$. This confirms magnetically the crystallographic observation that there is no mixing between Rh and Ti in the crystal structure, as such mixing in the dimer is expected to lead to a spin glass transition [40,41]. **Figure 3b** shows the magnetization as a function of applied magnetic field in the $\mu_0H$ = 0-9 T range at different temperatures. The M vs. H curves are linear above 10 K, and become curved below 10 K. There is no sign of magnetic saturation up to $\mu_0H$ = 9 T at temperatures down to 1.8 K.

The observation of magnetism in an Rh-based oxide is rare but not unique (see e.g. ref. 22). In $Ba_9Rh_8O_{24}$, for example, the fact that the magnetic susceptibility does not follow the Curie-Weiss law has been attributed to the presence of non-magnetic $Rh^{3+}$ ions that interrupt the exchange interactions between $Rh^{4+}$ ions, resulting the disruption in magnetic interactions within a chain[42]. $Sr_4RhO_6$, a magnetically ordered $Rh^{4+}$ compound[22], has an antiferromagnetic ordering transition at 7 K and an effective moment of 1.71 $\mu_B$/mol-Rh, as expected for moment of spin ½. Compared to $Sr_4RhO_6$, $Ba_6Y_2Rh_2Ti_2O_{17-\delta}$ has much smaller Curie-Weiss temperature and effective moment.

**Figure 4a** shows the heat capacity of $Ba_6Y_2Rh_2Ti_2O_{17-\delta}$ from 0.35 to 14 K measured in magnetic fields of $\mu_0H$ = 0, 1, 2, 4, 5, 6 and 9 T. A broad hump typical of the Schottky effect was observed at around 1 K under zero applied magnetic field[25]. This upturn is very sensitive to magnetic fields in the $\mu_0H$ = 1 Tesla range. A similar upturn feature at the lowest temperatures, suppressed by applied magnetic fields, has been observed in some QSL candidates[2,9,43,44]. The low temperature heat capacity approximately obeys a universal scaling law, as shown in **Figure 4b**[45]. Similar to reported QSL candidates, the heat capacity data of $Ba_6Y_2Rh_2Ti_2O_{17-\delta}$ collapse at the critical point of around 0.35 K/T[43,44,46–48]. The power law fitting of the zero applied field heat capacity data in $Ba_6Y_2Rh_2Ti_2O_{17-\delta}$ results in the power of 0.57 (**Figure 4c**). This is in contrast to what is seen in conventional magnetic insulators[26], whose heat capacity is proportional to $T^3$. We rule out the possibility that the nuclear spins of the constituent atoms are the origin of the sub-10-

Kelvin heat capacity behavior because the heat capacity of $Ba_6Y_2Ti_4O_{17}$ obeys $T^3$-behavior below 7 K and does not show the upturn (**Figure 4d**) and the heat capacity values in this temperature range for $TiO_2$, $Y_2O_3$, BaO and $RhO_2$ are many orders of magnitude smaller than that of $Ba_6Y_2Rh_2Ti_2O_{17-\delta}$ [49–52].

On subtraction of an estimated phonon contribution, the measured heat capacity data under zero applied magnetic field can be used to calculate the magnetic entropy in $Ba_6Y_2Rh_2Ti_2O_{17-\delta}$, as shown in **Figure 5a**. The heat capacity of $Ba_6Y_2Ti_4O_{17}$ from 0.35 to 12 K (where the heat capacity data for $Ba_6Y_2Rh_2Ti_2O_{17-\delta}$ and $Ba_6Y_2Ti_4O_{17}$ intersect) accounts for the phonon contribution. The magnetic entropy reaches only 0.32 Rln2 by 12 K. This suggests that a significant amount of magnetic entropy is present at temperatures below 0.35 K in this material. In the inset of **Figure 5a**, a plot of C/T as function of $T^2$ at zero applied magnetic field yields a zero temperature intercept of 166 mJ/mol f.u-$K^2$ (or 83 mJ/mol Rh-$K^2$). Compared to the superconductor $Sr_2RuO_4$ ($\gamma$ = 39 mJ/mol Ru-$K^2$)[53] or the $Sr_4Rh_3O_{10}$ layered perovskite ($\gamma$ = 16 mJ/mol Rh-$K^2$)[24] or $Sr_5Rh_4O_{12}$ ($\gamma$ = 30 mJ/mol-$K^2$)[39], $Ba_6Y_2Rh_2Ti_2O_{17-\delta}$, which is an electrical insulator, has a much larger linear contribution to its sub-Kelvin heat capacity. This may be of interest for future studies.

The direct band gap Tauc plot for $Ba_6Y_2Rh_2Ti_2O_{17-\delta}$ powder is shown in **Figure 5b**. Fitting to indirect band gap does not produce a reasonable, flat baseline, and thus the fit is done to the direct gap equation. The direct optical band gap is determined from this data to be approximately 0.17 eV, found by extrapolating the linear absorption region until it intersects with the baseline of the absorption as seen in **Figure 5b**. The hump in the absorption spectrum around 0.18 eV (1450 $cm^{-1}$) likely corresponds to a vibrational mode, supported by the observation of a vibrational mode at this energy in in the Raman spectrum (**Figure 5c**). The resistivity of $Ba_6Y_2Rh_2Ti_2O_{17-\delta}$ is plotted as a function of reciprocal temperature in the inset of **Figure 5b**. The activation energy of 0.15 eV, obtained by fitting the Arrhenius equation, which must be from or to defect states associated with the valence or conduction band, together with the optical data, show that $Ba_6Y_2Rh_2Ti_2O_{17-\delta}$ has a sub-eV band gap, and a high resistivity at low temperatures.

**CONCLUSIONS**

$Ba_6Y_2Rh_2Ti_2O_{17-\delta}$, a magnetic analog to the non-magnetic dielectric $Ba_6Y_2Ti_4O_{17}$, has been discovered and synthesized by a solid state method in air at ambient pressure. The material crystallizes in a hexagonal perovskite type structure in the *P6₃/mmc* space group. The $Rh_2O_9$ dimers present are significantly closer within the triangular planes than they are between planes, suggesting that the high temperature magnetism will be two-dimensional in character. The material has a small effective moment, 0.76 $\mu_B$/mol-Rh, or 1.5 $\mu_B$/dimer. The small negative Curie-Weiss temperature may be due to a competition between antiferromagnetic and ferromagnetic interactions, but it may alternatively indicate the dominance of very weak antiferromagnetic interactions in the material. The magnetic susceptibility and heat capacity measurements indicate that magnetic ordering is absent down to 0.35 K. The large T-linear contribution to the heat capacity at temperatures below 1 K may come from the presence of strong magnetic fluctuations. The heat capacity shows an upturn under zero field at the lowest temperatures measured that is suppressed by applied fields larger than 2 T, a characteristic that has been observed in other QSL candidates. The approximate data collapse in scaled heat capacity curves further suggests that $Ba_6Y_2Rh_2Ti_2O_{17-\delta}$ may host a spin liquid ground state in the rhodate perovskites for the first time. Thus growing crystals of this material will be of interest for further characterization to confirm the presence or absence of a spin liquid state.

**ACKNOWLEDGMENTS**

The authors acknowledge the use of Princeton's Imaging and Analysis Center, which is partially supported by the Princeton Center for Complex Materials, a National Science Foundation (NSF)-MRSEC program (DMR-1420541). All of the research reported here was supported by the Institute of Quantum Matter, an Energy Frontier Research Center funded by the U.S. Department of Energy, Office of Science, Basic Energy Sciences under Award No. DE-SC0019331. A portion of this research used resources at the Spallation Neutron Source, a DOE Office of Science User Facility operated by the Oak Ridge National Laboratory.

**Table 1**. Structural parameters for $Ba_6Y_2Rh_2Ti_2O_{17-\delta}$ at 300 K. Space group $P6_3/mmc$ (No. 194).

| Atom | Wyckoff. | Occ. | x | y | z | $U_{iso}$ |
|---|---|---|---|---|---|---|
| Ba1 | 4f | 1 | ⅓ | ⅔ | 0.1820(2) | 0.0022(2) |
| Ba2 | 4f | 1 | ⅓ | ⅔ | 0.5906(2) | 0.0014(2) |
| Ba3 | 2a | 1 | 0 | 0 | 0 | 0.036(3) |
| Ba4 | 2b | 1 | 0 | 0 | ¼ | 0.015(2) |
| Rh | 4f | 1 | ⅓ | ⅔ | 0.7073(3) | 0.073(3) |
| Ti | 4f | 1 | ⅓ | ⅔ | 0.0537(3) | 0.021(3) |
| Y | 4e | 1 | 0 | 0 | 0.1266(1) | 0.0045(1) |
| O1 | 12k | 1 | 0.172(2) | 0.344(2) | 0.42209(8) | 0.026(1) |
| O2 | 12k | 1 | 0.173(1) | 0.346(1) | 0.83111(7) | 0.011(1) |
| O3 | 6h | 1 | 0.520(2) | 0.040(2) | ¼ | 0.0058(2) |
| O4 | 4f | 0.81(2) | ⅓ | ⅔ | 0.5050(3) | 0.059(4) |

a = b = 5.9364(2) Å, c = 29.512(2) Å, V = 900.69(5) Å$^3$, α = β = 90°, γ = 120°.

wR = 6.50%, $R_F^2$ = 7.83%.

**Table 2**: Selected interatomic distances (Å) for $Ba_6Y_2Rh_2Ti_2O_{17-\delta}$ at 300 K.

|  | Interatomic distance (Å) |
|---|---|
| Rh-O2 (x3) | 1.999(5) |
| Rh-O3 (x3) | 1.966(5) |
| Ba1-O2 (x6) | 2.994(8) |
| Ba1-O3 (x3) | 2.778(4) |
| Ba2-O1 (x6) | 2.993(1) |
| Ba2-O2 (x3) | 2.837(5) |
| Ba2-O4 (x1) | 2.526(1) |
| Ba3-O1 (x6) | 2.8976(2) |
| Ba4-O2 (x6) | 2.9832(2) |
| Ba4-O3 (x6) | 2.975(1) |
| Ti-O1 (x3) | 1.811(4) |
| Ti-O4 (x1) | 1.732(1) |
| Y-O1 (x3) | 2.275(3) |
| Y-O2 (x3) | 2.174(3) |

**Figure Captions:**

**Figure 1**: (a) The SEM/EDX analysis of $Ba_6Y_2Rh_2Ti_2O_{17-\delta}$ confirms the metal elemental ratios. (b) The TGA analysis of $Ba_6Y_2Rh_2Ti_2O_{17-\delta}$ suggests the presence of oxygen vacancies. (c) and (d) Rietveld powder x-ray diffraction and neutron diffraction refinement of the crystal structure of $Ba_6Y_2Rh_2Ti_2O_{17-\delta}$ at 300 K in the space group $P6_3/mmc$ (No. 194). For neutron: wR = 6.50%, $R_F^2$ = 7.83%. For x-ray: wR = 11.18%, $R_F^2$ = 8.91%.

**Figure 2**: (a) Crystal structure of $Ba_6Y_2Rh_2Ti_2O_{17-\delta}$. (b) Triangular lattice of Rh magnetic dimers in the ab plane. (c) $Rh_2O_9$ dimer (two face-sharing $RhO_6$ octahedra) with one oxygen site depicted in red and the other orange. There are two different Rh-O bond lengths, 2.00 Å and 1.97 Å. The Rh-Rh distance within the dimer is 2.52 Å.

**Figure 3**: (a) Temperature dependent magnetic susceptibility in $Ba_6Y_2Rh_2Ti_2O_{17-\delta}$ under an applied field of 2 kOe. Inset: Curie-Weiss fitting from 10-50 K results in the effective moment of 0.76 $\mu_B$/mol-Rh and the Curie-Weiss temperature of -2.8 K. ZFC/FC magnetic susceptibility of $Ba_6Y_2Rh_2Ti_2O_{17-\delta}$ under an applied magnetic field of 100 Oe. The absence of bifurcation between ZFC (black squares) and FC (red circles) rules out the possibility of spin glass behavior and atomic disorder in this material. (b) Magnetization as a function of applied magnetic field from 0-9 T at different temperatures. There is no sign of magnetic saturation up to 9 T at temperatures down to 1.8 K.

**Figure 4**: (a) Heat capacity divided by temperature for $Ba_6Y_2Rh_2Ti_2O_{17-\delta}$ measured under different applied magnetic fields below 15 K. At zero field, the presence of a high density of low energy states is evidenced by the presence of a large upturn that persists down to the lowest temperature of the measurement. The small hump at 1 K may come from the Schottky effect, where the spins are populated from one to another energy level. (b) The universal scaling of heat capacity data collapsed at the critical point[45]. (c) The power law fitting of the raw heat capacity data in $Ba_6Y_2Rh_2Ti_2O_{17-\delta}$ results in the power of 0.57. Inset: The existence of heavy quasiparticles in insulating $Ba_6Y_2Rh_2Ti_2O_{17-\delta}$ is supported by a very large Sommerfeld constant [$\gamma$ = 166 mJ $mol_{f.u}^{-1}$ $K^{-2}$]. (d) We rule out the possibility that the observed behavior arises from the nuclear spins of the constituent atoms by measuring the heat capacity of $Ba_6Y_2Ti_4O_{17}$ (no upturn observed). Moreover,

the nuclear spins in $TiO_2$, $Y_2O_3$, $BaO$ and $RhO_2$ result in a low temperature heat capacity orders of magnitude smaller than the value of 700 mJ/mol-$K^2$ at 0.35 K seen in $Ba_6Y_2Rh_2Ti_2O_{17-\delta}$.

**Figure 5**: (a) C/T vs. T in $Ba_6Y_2Rh_2Ti_2O_{17-\delta}$ measured down to 0.35 K under zero applied magnetic field. Below 1 K, the upturn indicates the presence of spin fluctuations. The heat capacity of $Ba_6Y_2Ti_4O_{17}$ below 12 K was used to model the phonon contribution. The magnetic entropy is calculated to be 1.86 J/mol$^{-1}$K$^{-1}$, accounting for about 1/3 of the magnetic entropy of a two state magnetic system. (b) Direct band gap Tauc plot of $Ba_6Y_2Rh_2Ti_2O_{17-\delta}$ with optical band gap of 0.17 eV. Inset: Resistivity measurement on $Ba_6Y_2Rh_2Ti_2O_{17-\delta}$ resulting in a transport gap of 0.15 eV. (c) Raman scattering spectrum of $Ba_6Y_2Rh_2Ti_2O_{17-\delta}$.

**Figures:**

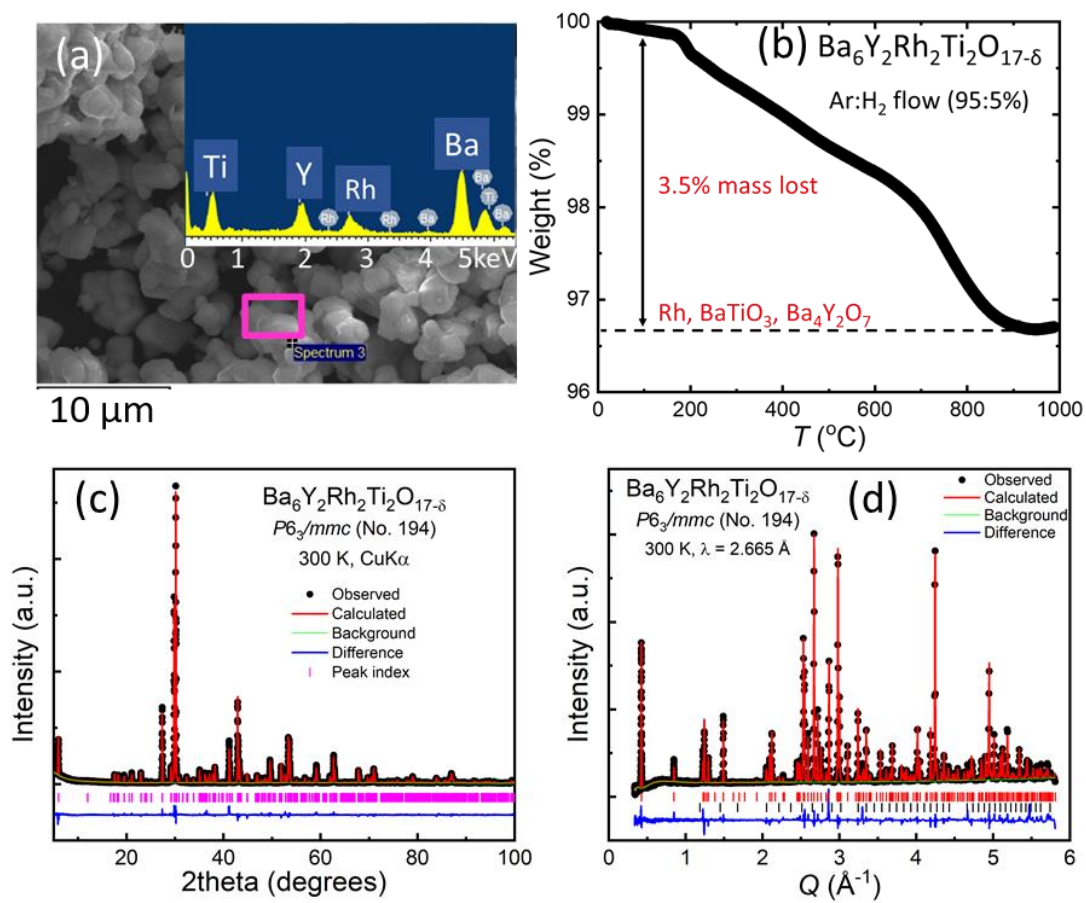

**Figure 1**.

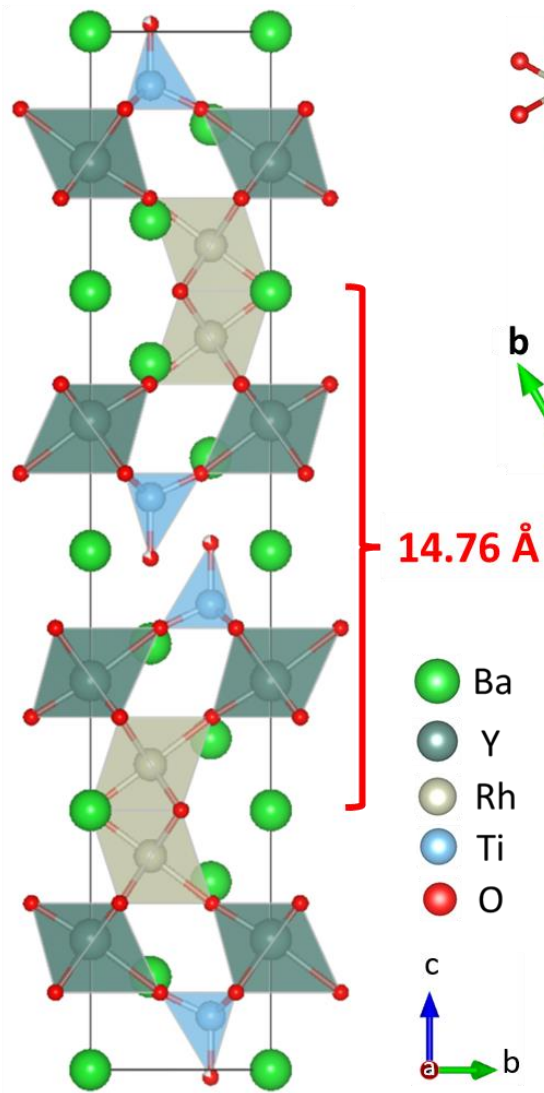
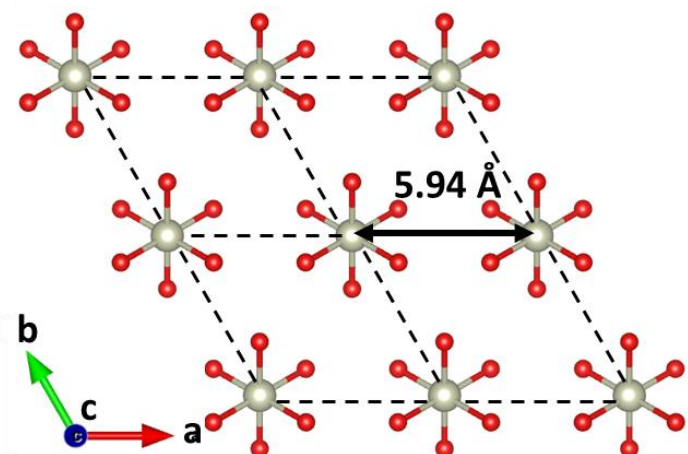
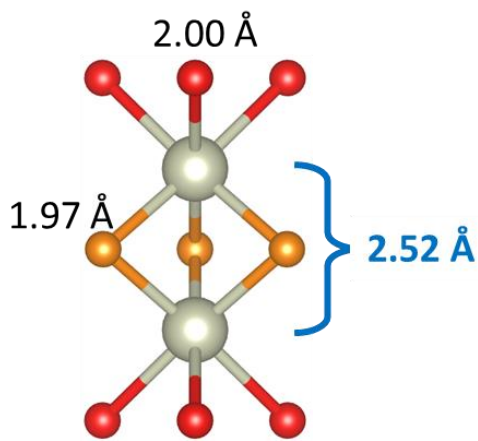

**Figure 2**.

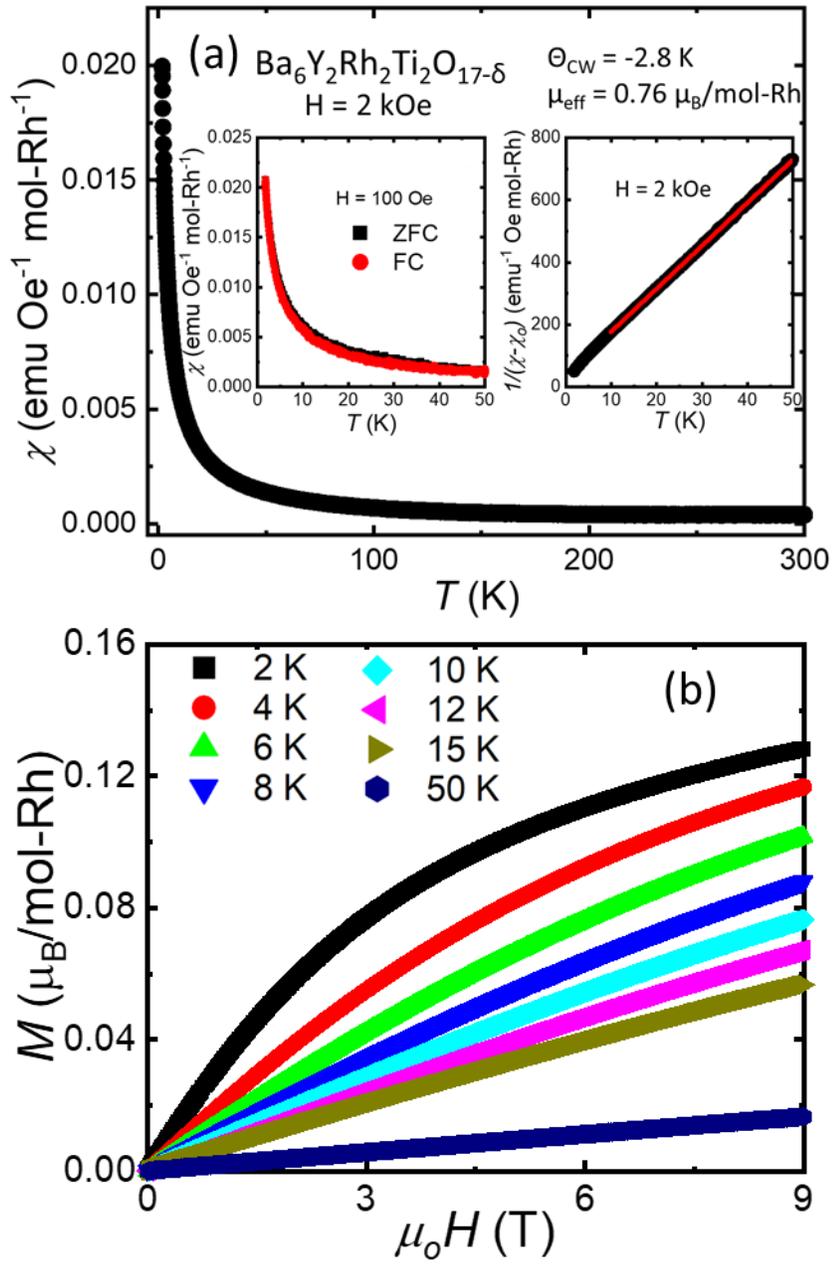

**Figure 3**.

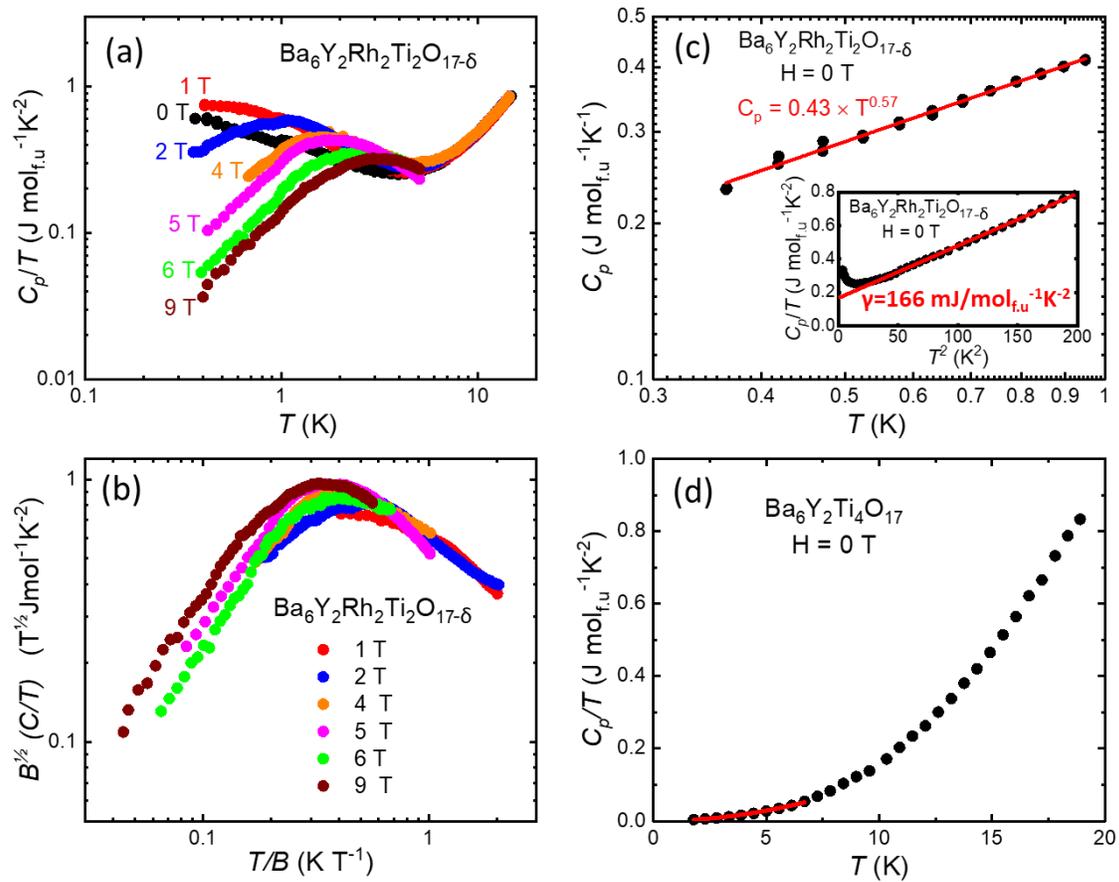

**Figure 4**.

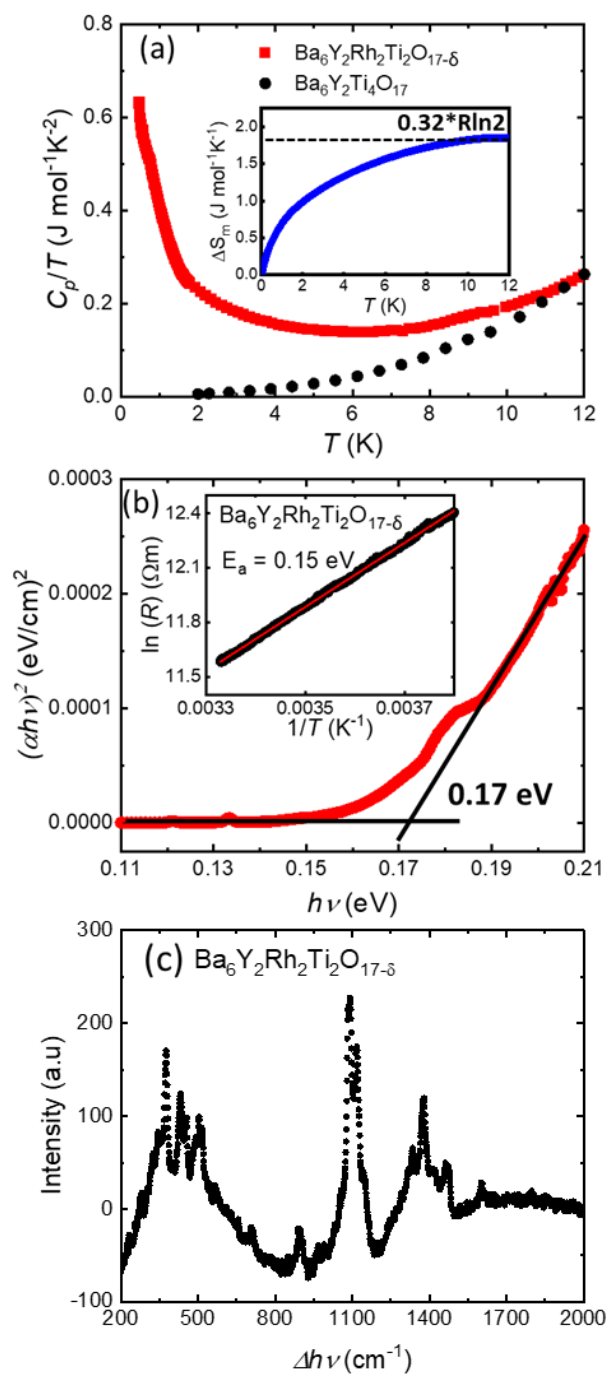

**Figure 5**.